\documentclass{article}

\usepackage{arxiv}

\usepackage[utf8]{inputenc} 
\usepackage[T1]{fontenc}    
\usepackage{hyperref}       
\usepackage{url}            
\usepackage{booktabs}       
\usepackage{amsfonts}       
\usepackage{nicefrac}       
\usepackage{microtype}      
\usepackage{cleveref}       
\usepackage{lipsum}         
\usepackage{graphicx}
\usepackage{doi}
\usepackage{fancyvrb}
\usepackage{enumitem}

\usepackage{multicol}

\usepackage{tabularx}
\usepackage{capt-of}

\title{Reproducing Random Forest Efficacy in Detecting Port Scanning}

\author{
    Jason M. Pittman \\
    ORCID: 0000-0002-5198-8157 \\
}

\hypersetup{
    pdftitle={Reproducing Random Forest Efficacy in Detecting Port Scanning}
    pdfsubject={}
    pdfauthor={Jason M. Pittman}
    pdfkeywords={reproduction, random forest, machine learning, port scanning, cybersecurity, algorithms}
}

\begin{document}
\maketitle

\begin{abstract}
Port scanning is the process of attempting to connect to various network ports on a computing endpoint to determine which ports are open and which services are running on them. It is a common method used by hackers to identify vulnerabilities in a network or system. By determining which ports are open, an attacker can identify which services and applications are running on a device and potentially exploit any known vulnerabilities in those services. Consequently, it is important to detect port scanning because it is often the first step in a cyber attack. By identifying port scanning attempts, cybersecurity professionals can take proactive measures to protect the systems and networks before an attacker has a chance to exploit any vulnerabilities. Against this background, researchers have worked for over a decade to develop robust methods to detect port scanning. One such method revealed by a recent systematic review is the random forest supervised machine learning algorithm. The review revealed six existing studies using random forest since 2021. Unfortunately, those studies each exhibit different results, do not all use the same training and testing dataset, and only two include source code. Accordingly, the goal of this work was to reproduce the six random forest studies while addressing the apparent shortcomings. The outcomes are significant for researchers looking to explore random forest to detect port scanning and for practitioners interested in reliable technology to detect the early stages of cyber attack.   
\end{abstract}


\begin{multicols}{2}

\section{Introduction}
Cybersecurity breaches are a persistent threat in the digital world.. Although phishing and malware account for a considerable number of cybersecurity incidents, 45\% are caused by network-based cyber attacks \cite{statistia}. Cyber attacks follow a set pattern or sequence of steps, although the number of steps varies among existing models and methodologies. Nonetheless, reconnaissance is universally recognized as the first step in these attacks. Reconnaissance typically involves performing some form of port scanning.

Port scanning is a method of identifying active services on a target system. The technique can be both a legitimate means to verify one's configurations \cite{Wang2014} or as a precursor to an intrusion \cite{Baig2007, yadav2015technical}. Thus, a general problem is differentiating between what may be an authorized  host enumeration versus a malicious scan. Moreover, if port scanning is a standard prelude to cyber attack, then we want to have a way to detect the technique with high certainty. To this end, there is a literature detailing a variety of port scan detection mechanisms. The literature covers intrusion detection systems relying on static signatures \cite{heberlein1990network} to advanced, modern machine learning (ML) techniques \cite{henry2023composition}. 

Most recently, a systematic review \cite{pittman2023machine} collected the body of work using ML to detect port scanning published since 2021. The review found 15 studies exploring five different algorithms. Those 15 studies demonstrated algorithm efficacy ranging from 37.92\% to 100\%. Based on the span of the algorithm efficacy, the review suggested follow-up replication of those studies be performed to determine why algorithm performance results varied so much across the literature. However, replication of all five algorithms is infeasible when accounting for the variations in hyperparameters, model training mechanisms, lack of available code, and so forth. Accounting for such is necessary because the majority (13 of 15) studies do not include or link to source code containing these implementation details. Therefore, we choose to focus on the random forest algorithm in this work because of its simple architecture but robust set of options.  

The remainder of this work is organized in a way which (a) situates the reproduction of random forest research in existing knowledge and (b) maximizes understanding of the cutting edge. The first is achieved by discussing port scanning, the random forest algorithm, and port scan detection using random forest implementations. Thereafter, we discuss the research method and protocol used to generate and analyze data. Finally, we present the results of the analysis.
 
\section{Related Work}
The work most proximal to this study exists in two categories: scanning TCP/IP ports, random forest in ML, and the use of random forest to detect port scans. The following background discussion is not meant to be comprehensive, however. Rather, we present notable and relevant research. The intention is to construct a suitable conceptual framework for understanding how and why this work has significance to the field. 

\subsection{Port Scanning}
The origin of the term \textit{port scanning} in the academic literature can be traced back to the late 1980s and early 1990s as the Internet was growing and becoming more widely used. Port scanning involves using features of TCP/IP to gather information about computing systems on a network by identifying open or closed ports. The first reference to port scanning in the literature is found in the work of Fyodor \cite{lyon1998} which described a method for determining the operating system of a remote host by sending probes to specific ports and analyzing the responses. This paper also introduced the first version of the open-source tool \textit{nmap}. Later, De Vivo et al. \cite{de1999review} provided a classification of port scanning techniques and procedures into different types such as TCP connect scans, SYN scans, indirect scans, stealth scans, and others. Later, Barnett et al. \cite{barnett2008towards} presented a classification system for network scanning techniques based on the level of interaction with the target system, the type of information gathered, and the purpose of the scan. Bou et al. \cite{bou2013cyber} demonstrated a comprehensive overview of different types of cyber scanning techniques that are used to identify various features of networks, which can be divided into passive and active scanning techniques.

\subsection{Random Forest}
Random forest is a supervised ensemble ML algorithm that leverages a set (i.e., forest) of decision trees to make a prediction \cite{breiman2001random}. The name random forest refers to the fact that the each tree in the forest is trained on a random subset of the data and a random subset of the features. A final prediction is made by combining the predictions of all the trees in the forest, either by majority vote (for classification problems) or by averaging (for regression problems). While random forest can act as a regression tool and a classifier, the algorithm is particularly robust for classification problems \cite{breiman2001random, schonlau2020random, xu2012improved, vcehovin2010empirical}. This is true particularly in cases where there are many features and interactions among features.

While the literature exhibits a variety of differing random forest algorithm implementations, all consist of four general steps:
\begin{itemize}[leftmargin=*]
	\item Select random samples from a given dataset.
	\item Construct a decision tree for each sample and get a prediction result from each decision tree.
	\item Perform a vote for each predicted result.
	\item Select the prediction result with the most votes as the final prediction.
\end{itemize}

\subsubsection{Hyperparameters and Parameters}
Hyperparameters and parameters are not synonymous. Most classification algorithms use hyperparameters. Hyperparameters are set before training a model and cannot be estimated from the data during training \cite{breiman2001random, schonlau2020random}. The hyperparameters control the behavior and performance of the model as well as optimization. Conversely, parameters are estimated from the data during training of a machine learning model. These values are learned by the model during optimization and are input to generate predictions.  

To that end, the set of common hyperparameters one can manipulate when using random forest includes:

\begin{itemize}[leftmargin=*]
	\item Estimators as the number of decision trees in the forest.
    	\item Criterion as the measure used to split the tree at each node.
	\item The maximum depth of the decision trees.
	\item A minimum number of samples required to split an internal node.
    	\item The minimum number of samples required to construct a leaf node.
    	\item The number of features to consider when looking for the best split.
    	\item Whether or not to sample with replacement.
\end{itemize}

A crucial hyperparameter to adjust is the number of random features to be taken into account at each split point. Another important aspect to adjust is the depth of the decision trees \cite{schonlau2020random}. Although deeper trees tend to overfit the training data, they also exhibit lower correlation, which can potentially enhance the performance of the ensemble. Nonetheless, node size, tree quantity, and number of sampled features are generally accepted as mandatory to set \cite{fawagreh2014random, kumar2019advantages}.

Still, even with perfect hyperparameters controlling training, random forest has advantages and disadvantages. Awareness of both can lead to more appropriate selection of the algorithm for a given problem. The same is true at the level of how the algorithm is configured through hyperparameters. 

\subsubsection{Advantages of random forest}
The main advantage of Random Forest classifiers is that they are less likely to overfit than single decision trees, since the combination of multiple trees helps to average out the noise in the data \cite{fawagreh2014random, kumar2019advantages}. Random forest is also relatively easy to implement and use. This is because the algorithm produces trained models requiring little tuning compared to other machine learning algorithms \cite{breiman2001random, schonlau2020random}. 

Another significant advantage of random forest is it is an ensemble method \cite{schonlau2020random}. Other classifiers like logistic regression, support vector machine, and naive Bayes are single models. Random forest also uses the bagging technique to train each tree in the forest on a random subset of the data and a random subset of the features \cite{breiman2001random, kumar2019advantages, schonlau2020random}. This helps to reduce the risk of overfitting and improve the stability of the model. As well, this allows random forest to compute feature importance, which is a measure of the contribution of each feature to the prediction \cite{breiman2001random, schonlau2020random}. This can be useful for feature selection and interpretability.

\subsubsection{Disadvantages of random forest}
In terms of disadvantages, random forest classifiers may not perform as well on highly complex and non-linear problems \cite{fawagreh2014random, kumar2019advantages}, where other algorithms like neural networks may be more suitable. On that note, random forest classifiers can be computationally expensive and slow to make predictions, especially for large datasets and many trees in the forest \cite{schonlau2020random}. Further, random forest classifiers may not handle unbalanced data well, where one class is underrepresented \cite{breiman2001random, kumar2019advantages, schonlau2020random}. This can lead to biased results towards the majority class. Finally, while random forest classifiers can provide feature importance such are not as interpretable as other models like linear regression  where the coefficients have a clear meaning.

Regardless of advantages and disadvantages, random forest features prominently in port scan detection research. While collective results of the research since 2021 demonstrated high efficacy, the deviation of accuracy values appeared significant. The following section describes those values along with other pertinent random forest implementation details. 

\subsection{Detecting Port Scanning}
Port scanning, as a reconnaissance technique, is detectable. ML is a compelling solution to detecting otherwise undetectable port scans because of its ability to correlate seemingly unrelated features across enormous datasets. Yet, not all ML algorithms work in the same way or have the capability to address the same problem. 

Furthermore, there are a variety of ML algorithms types- classification, regression, deep learning, and so forth- with a diversity of implementation variations. This work focused on random forest and therefore we analyzed the following background works only for related details to that algorithm.

\begin{center}
\captionof{table}{Random Forest Algorithm Literature}
\begin{tabular}{cccc}
	\textbf{Authors} & \textbf{Accuracy} & \textbf{Dataset} \\ \hline \hline
	Algaolahi et al. & 99.75 & CICIDS2017\\
	Baah et al. & 99.98 & CICIDS2017\\
	Sirisha et al. & 78.09 & NSLKDD \\
	Sirisha et al. & 84.14 & CICIDS2017 \\
	SaiKiran et al. & 99.93 & CICIDS2017 \\
	Mohseni et al. & 99.94 & CICIDS2017 \\
	Bertoli et al. & 96.00 & MAWILab \\ 
	Bertoli et al. & 100.00 & Bonafide \\ \hline
\end{tabular}
\end{center}

The six studies \cite{algaolahi2021port, bertoli2021end, sirisha2021intrusion, mohseni2021density, baah2022enhancing,  saikira2022detection} in Table 1 experimented with port scan detection using different RF algorithm implementations. Overall, random forest performance ranged from 78.09\% to 100\% across those studies. One paper \cite{bertoli2021end} included a link to source code in a public repository (e.g., GitHub). Four different datasets were used, three of which do not appear in other algorithm categories.

Two studies discussed the types of port scans present in training and evaluation data. SaiKiran et al. \cite{saikira2022detection} mentioned \textit{port sweep} but did not specify further. Bertoli et al. \cite{bertoli2021end} conducted training and evaluating against the full spectrum of port scan types. Further, Bertoli et al. included port scan data from five different port scan tools.

Algaolahi et al. \cite{algaolahi2021port} used a 20\% test, 80\% training data split. Random forest hyperparameters were not provided. Baah et al. \cite{baah2022enhancing} also used a 20-80\% data split. Additionally, the authors specified a 10-fold cross-validation value for training. Other hyperparameters were absent. 

Meanwhile, Bertoli et al. \cite{bertoli2021end} provided perhaps the most detailed description of their experimental configuration. Random forest grid-search was performed with estimators set to 10, 50, 100, and 200. The criteria were gini and entropy. A max depth of 5 and 10. Lastly, the authors set a class weight of \textit{none, balanced, and balanced subsample}. The model evaluation used stratified k-fold (k=10) cross-validation. 

Mohseni et al. \cite{mohseni2021density} also described a set of hyperparameters for their implementation of the random forest algorithm. However, details were unclear. It is possible the class weight was balanced. Further, it seemed the estimators were set to 15, 25, 50, and 100. Training data split and cross-validation were not mentioned. 

Less revealing, SaiKiran et al. \cite{saikira2022detection} did not indicate any random forest hyperparameters, training data split, or cross-validation mechanism. Sirisha et al. \cite{sirisha2021intrusion} provided some hyperparameters such criterion as gini, estimator of 100, but nothing else. The authors used a data split of 25\% test and 75\% training.

\section{Method}
A recent systematic review \cite{pittman2023machine} suggested follow-up replication be performed for research using ML to detect port scanning. Research replication and reproduction is often mentioned but seldom implemented. For completeness, replication duplicates a research study with the same methods and setup to verify and increase trust in results \cite{plesser2018reproducibility}. The feasibility of conducting replication depends on the degree to which instrumentation and experimental setups are discussed in the source studies \cite{lindsay1993design, brooks1996replication, gomez2010replications}. Reproduction is related but different insofar as it recreates a study using different instruments or data sets to demonstrate generalizability in the source studies' results \cite{plesser2018reproducibility}. Reproduction does not depend on instrument or experimental setup details \cite{lindsay1993design, brooks1996replication, gomez2010replications}.

However, as noted both in the systematic review and the related work in this research, the body of random forest port scan detection literature lacks the necessary details to establish a replication. Thus, a reproduction is the most appropriate method \cite{plesser2018reproducibility}. 

Scientific reproduction is meaningful when such work follows the source literature as closely as possible, is fully transparent in its method and materials, and rigorously compares reproduction results to the source research. Thus, we pose a single research question to guide the reproduction process. The question is: to what extent does reproduction produce similar results to the original studies, and if not are the differences statistically significant? To generate an answer we measured the efficacy of the random forest algorithm under several implementations. 

\subsection{Protocol}
With the above in mind, we constructed a reproduction protocol based on (a) extracted details from the source studies where available and (b) knowledge of how random forest can be implemented. Having a clear protocol ensured we handled important steps in the correct sequence. Moreover, a clear protocol is a key tenet in reproduction, especially when source work lacks clarity.

\begin{enumerate}[leftmargin=*]
	\item Initialize the environment
	\item Select a dataset for random forest training and testing
	\item Conduct necessary data pre-processing
	\item Run a random forest trial for hyperparameter set $A$
	\item Record the \Verb|score| output
	\item Re-initialize the environment
	\item Run a random forest trial for hyperparameter set $B$
	\item Record the \Verb|score| output
	\item Repeat steps 6 - 8 until all hyperparameter sets are trialed
\end{enumerate}

\subsection{Environment} 
We constructed the operational environment in two stages: hardware and software. The hardware consisted of an AMD Ryzen 9 5900X CPU, 128 gigabytes of RAM, and a Nvidia 3090 RTX GPU. The system ran Ubuntu 22.04 for an operating system. The ML stack included \Verb|Python 3.11.1|, \Verb|Numpy 1.24.1|, \Verb|Pandas 1.5.3|, and \Verb|Scikit-Learn 1.2.1|.   

\subsection{Datasets}
The source studies used four different datasets. The CICIDS2017 dataset includes eight different files, one being explicitly labeled as having \textit{port scan} material (Friday-WorkingHours-Afternoon-PortScan.pcap\_ISCX). None of the studies using this dataset detail which of the eight files were used for training and testing. Further, recent work  \cite{lanvin2023errors} discovered significant issues with the CICIDS2017 set including duplicate data and mislabeling. The NSLKDD dataset also includes multiple files. Sirisha et al. \cite{sirisha2021intrusion} used the KDDTrain+ and KDDTest+ files. However, Bertoli et al. \cite{bertoli2021end} noted limitations of the data which rendered use of the data in ML-based solutions impractical. 

In turn, Bertoli et al. used two different datasets. One dataset contained laboratory generated traffic (attack) while the other the contained traffic captured from an internet network segment (bonafide). We selected the bonafide dataset because it included both legitimate as well as port scan traffic and therefore more closely matched the contents of the CICISD2017 and NSLKDD datasets. Moreover, we manually inspected the bonafide dataset to ensure a variety of port scan tools and techniques were included. For tools, the dataset captured nmap, masscan, unicornscan, zmap, and hping.  

We adhered to the data pre-processing process outlined by Bertoli et al. \cite{bertoli2021end} with one deviation. We opted to save the final processed dataset to comma separated value file. Doing so saved time and reduced error surface given we were going to run multiple random forest trials. The dataset is available in the companion GitHub repository.

\subsection{Hyperparameters}
There are five parameters within the set of hyperparameters that can impact random forest accuracy based on Bertoli et al. \cite{bertoli2021end} study implementation. We gave critical attention to three as we constructed the sets for reproduction trials. First, we let the two lists of \Verb|n_estimators| from source studies be hyperparameter sets $A$ and $B$. This parameter defines the number of trees in the model. Then, we paired value definitions of \Verb|max_depth| also from source work to those sets. Here, the deeper the tree, the more splits can occur causing the model to capture more information about the data. Lastly, we isolated \Verb|min_samples_leaf| as the number of samples required to be at a leaf node. No source work set this parameter so we let the default value hold for sets $A$ and $B$. 

The majority of source studies (4 of 6) did not specify hyperparameters. Thus, we took guidance from the literature to construct additional hyperparameter sets $C$ and $D$. We used these sets as internal validity measures and for comparison of model efficacy under different random forest training conditions. Additionally, the sets $C$ and $D$ allow for comparison to the broader literature not reporting explicit implementation details. 

\begin{center}
\captionof{table}{Hyperparameter sets and parameter values}
\begin{tabular}{ccccccc}
\textbf{Set} & \textbf{Estimators} & \textbf{Leaf}  & \textbf{Depth} \\ \hline \hline
A   & 10, 50, 100, 200 & 1	& 5, 10 \\ 
B   & 15, 25, 50, 100	& 1 & 5, 10 \\ 
C   & 200, 500 & 14 & 4, 5, 6, 7, 8 \\ 
D   & 200, 500 & 14 & 4, 5, 6, 7, 8 \\ \hline \hline 
\textbf{Set} & \textbf{Features} & \textbf{Criterion} & \textbf{Weight} \\
A & sqrt & gini & balanced \\
B & sqrt & gini & balanced \\
C & sqrt & gini & balanced \\
D & log2 & entropy & None \\ \hline
\end{tabular}
\end{center}

\subsection{Instrumentation}
We initially attempted to use the source code provided through Bertoli et al \cite{bertoli2021end}. However, there were critical errors during runtime. We surmised these were due to version drift in core packages (e.g., \Verb|sklearn|). As well, because none of the other source studies offered code snippets or entire source, we opted to develop a standard random forest implementation based on available \Verb|Scikit-Learn| documentation. Further, in doing so we establish a base implementation for a known working algorithm and model. Further, doing so allowed us to modularize the code within a Jupyter notebook according to the protocol rather than modify an existing instrument.

We let the training and testing split be 70\% and 30\% across all reproduction trials. This seemed appropriate given the known configurations in the source studies. Further, each hyperparameter set included two tuning configuration methods: \Verb|RandomSearchCV| and \Verb|GridSearchCV|. Considering the majority of source studies did not include code or indicate which method was used, including both provides a comparative view for later analyses. Fit and scoring were achieved using the basic random forest methods.

We ran an internal pilot test to ensure the instrument functioned as expected. This included loading the data, pre-processing the data, instantiating the random forest algorithm with a hyperparameter set, and producing output. Lastly, we made the instrumentation and data used in this reproduction transparent. All related materials can be found in a public GitHub repository \cite{pittman2023repo}.

\section{Results}
We separate the results into four sections based on the set of random forest configurations. Each results table contains two rows. The first row demonstrates results associated with the \Verb|RandomSearchCV| hyperparameter set and the second shows results using \Verb|GridSearchCV|. 

The first hyperparameter trial demonstrated macro averages of 99\%. These results can be taken as a baseline because they closely align to the random forest implementation used in Bertoli et al. \cite{bertoli2021end}.  

\begin{center}
\captionof{table}{Hyperparameter set A efficacy}
\begin{tabular}{cccc}
\textbf{Accuracy} & \textbf{Recall} & \textbf{Precision} & \textbf{F1} \\ \hline \hline
0.9970 & 0.99 & 0.99 & 0.99 \\
0.9976 & 0.99 & 0.99 & 0.99 \\ \hline
\end{tabular}
\end{center}

Changing the \Verb|n_estimators| to $[15, 25, 50, 100]$ produced the output in Table 4. While the macro averages remain at 99\%, we observed a minor difference in the \Verb|GridSearchCV| results compared to Set A.

\begin{center}
\captionof{table}{Hyperparameter set B efficacy}
\begin{tabular}{cccc}
\textbf{Accuracy} & \textbf{Recall} & \textbf{Precision} & \textbf{F1} \\ \hline \hline
0.9970 & 0.99 & 0.99 & 0.99 \\ 
0.9974 & 0.99 & 0.99 & 0.99 \\ \hline
\end{tabular}
\end{center}

We changed the \Verb|n_estimators| for Set C as well, to 200 and 500. We also changed \Verb|max_depth| from $[5, 10]$ used in the prior two trials to $[4, 5, 6, 7, 8]$. As well, we moved from the default \Verb|min_samples_leaf| value of $1$ to $14$. Collectively, these tunings produced changes to accuracy but not recall, precision, or F1 (Table 5).

\begin{center}
\captionof{table}{Hyperparameter set C efficacy}
\begin{tabular}{cccc}
\textbf{Accuracy} & \textbf{Recall} & \textbf{Precision} & \textbf{F1} \\ \hline \hline
0.9939 & 0.99 & 0.99 & 0.99 \\
0.9937 & 0.99 & 0.99 & 0.99 \\ \hline
\end{tabular}
\end{center}

We modified the last hyperparameter set with \Verb|max_features| from sqrt to log2 and a \Verb|criterion| value from gini to entropy. We also let \Verb|class_weight| be None. We observed changes in efficacy across all four measures.

\begin{center}
\captionof{table}{Hyperparameter set D efficacy}
\begin{tabular}{cccc}
\textbf{Accuracy} & \textbf{Recall} & \textbf{Precision} & \textbf{F1} \\ \hline \hline
0.9947 & 0.98 & 0.99 & 0.98 \\
0.9895 & 0.96 & 0.99 & 0.97 \\ \hline
\end{tabular}
\end{center}

As a general reproduction of random forest efficacy in detecting port scans, we were interested in how the above results compare to the results from the source literature. To that end, we expanded Table 1 by adding results for Recall, Precision, and F1 when available. The $NaN$ value indicates no data were available in that category from the source study.

\begin{center}
\captionof{table}{Efficacy comparison across source studies}
\begin{tabular}{lcccc}
\textbf{Study} & \textbf{Accuracy} & \textbf{Recall} & \textbf{Precision} & \textbf{F1} \\ \hline \hline
Algaolahi & 0.9975 & 0.9989 & 0.9975 & 0.9982 \\
Baah & 0.9998 & 0.9997 & 0.9999 & 0.9998 \\
Sirisha & 0.7650 & 0.6525 & 0.9721 & 0.7809 \\
SaiKiran & 0.9993 & NaN & NaN & NaN \\
Mohseni & 0.9964 & NaN & NaN & NaN  \\
Bertoli & NaN & NaN & NaN & 1.0000 \\ \hline
\end{tabular}
\end{center}

We statistically compared the \Verb|accuracy| results from our trials to the set of results from the source literature. We used a paired t-test and found no significant difference ($p = 0.4103$, $t = 0.9544$, $df = 3$, $std = 0.058$).

\section{Conclusion}
Port scanning is used by hackers to identify vulnerabilities in a network or system. Early detection of this step in the cyber kill chain can lend a sizable time advantage to network defense teams. Thus, much effort has been invested into developing techniques to do so. In that regard, researchers have put forth a variety of ML algorithms as solutions. Results from that work have shown promising results. However, whether such outcomes are generalizable has been unknown until now.

This study reproduced results from six studies published since 2021 using a random forest algorithm. We developed a new random forest instrument and ran it against one of the datasets detailed in one of the existing works. Further, to model potential variations in random forest implementation, we ran four trials with four different sets of hyperparameters. 

Given the observable similarity of results, combined with the statistical comparison, the use of random forest to detect port scanning appears generalizable. However, a deeper inspection of the trial data revealed that two techniques- full connect and SYN scanning- from one of the tools (unicornscan) were detected with an \Verb|accuracy| well below 90\%. 

We assumed the source research implemented the random forest algorithm within a common, standard range of hyperparameters. We also assumed given all studies used comparable Python interpreters and core packages given the work occurred within the same timeframe. However, it is possible major version differences exist across the studies or abnormal hyperparameters were used. We also assumed the selected dataset from Bertoli et al. \cite{bertoli2021end} did not contain duplicates, flaws, or mislabeled elements. 

Based on the results, we recommend additional reproduction research be done with focus on the other ML algorithms. Doing so would expand the foundation of knowledge and provide research and practitioners with confidence in the applicability of using ML to detect port scans. Furthermore, development of a real-time engine using confirmed generalizable ML algorithms would allow for experimentation with dynamic and distributed port scanning techniques. A component in such an engine might also give real-time measurements of sustainability or green compute impact. Lastly, deeper exploration of why unicornscan deviated from the high watermark predictions associated with other tools is warranted. Such investigation may explore novel port scanning techniques capable of evading ML-based solutions.

\bibliographystyle{unsrt}
\bibliography{references}

\end{multicols}
\end{document}